\documentclass[prl,twocolumn,amsmath,amssymb,showpacs,floatfix,nofootinbib,preprintnumbers]{revtex4-1}
\usepackage[latin1]{inputenc}
\usepackage{amsmath}
\usepackage{amsfonts}
\usepackage{amssymb}
\usepackage{graphicx}
\usepackage{dcolumn}
\usepackage{caption}
\usepackage{subfigure}
\usepackage{float}
\usepackage{bm}
\usepackage{latexsym}
\usepackage{epsfig}
\usepackage[normalem]{ulem}
\usepackage{graphicx}
\usepackage[colorlinks=true]{hyperref}
\begin{document}
\title{Litmus Test for Cosmic Hemispherical Asymmetry in the Cosmic Microwave Background B-mode polarization}
\author{Suvodip Mukherjee}\email{suvodip@iucaa.in}
\author{Tarun Souradeep}\email{tarun@iucaa.in}
\affiliation{Inter University Centre for Astronomy and Astrophysics \\ Post Bag 4, Ganeshkhind, Pune-411007, India}
\date{\today}
\begin{abstract}
Recent measurements of the temperature field of Cosmic Microwave Background (CMB) provide tantalising evidence for violation of Statistical Isotropy (SI) that constitutes a fundamental tenet of contemporary cosmology.  CMB space based missions, WMAP and Planck have observed a $7\%$ departure in the SI temperature field at large angular scales. However, due to higher cosmic variance at low multipoles, the significance of this measurement is not expected to improve from any future CMB temperature measurements. 
We demonstrate that weak lensing of the CMB due to scalar perturbations produce a corresponding SI violation in $B$ modes of CMB polarization at smaller angular scales. Measurability of this phenomenon depends upon the scales ($l$ range) over which power asymmetry is present. Power asymmetry which is restricted only to $l<64$ in temperature field cannot lead to any significant observable effect from this new window. However, this effect can put an independent bound on the spatial range of scales of hemispherical asymmetry present in scalar sector.
\end{abstract}
\pacs{98.70.Vc, 98.80.-k, 98.80.Qc}
\maketitle
By and large, the measurements of Cosmic Microwave Background (CMB) temperature and polarization are well explained by the LCDM model of cosmology. However, the evidence of Cosmic Hemispherical Asymmetry (CHA) in CMB temperature field, as observed by  both WMAP \cite{erikson} and Planck\cite{planck23, planckis,planck-oth}, has remained an  enigma for more than a decade. Several proposed origin of the CHA invoke different phenomena during inflation \cite{ha_ref_1} that generate characteristic signatures in both scalar and tensor perturbations. Hence, if CHA is indeed primordial, its nature and origin can be discerned by measuring it from CMB polarization field. However, due to higher cosmic variance and relative level of foreground contaminations at large angular scales, the direct appraisal of CHA from temperature as well as the polarization field at those scales is exigent.

In this Letter, we indicate an inevitable new avenue arising due to weak lensing of CMB photons by Statistical Isotropy (SI) violated Large Scale Structures (LSS), which  engenders unique signature of CHA in $B$-mode polarization at small angular scales. Hence this phenomenon opens a new window to measure CHA present in scalar perturbations and may validate the observation made from temperature field. The other known effect which produces isotropy violation in $B$ modes at small scales is due to our local motion \cite{doppler_mukherjee}. But it has negligible contribution in comparison to CHA.

The dominant contribution to $B$-mode polarization power spectrum at small angular scales arises from scalar perturbations due to weak lensing by the intervening LSS \cite{wayne, hu_2, lewis, marc_wl_1}. This inevitable mechanism that transfers power from $E$-mode to $B$-mode via the lensing kernel \cite{wayne, hu_2, lewis, marc_wl_1} is completely determined by the power spectrum of the observed LSS. Presence of SI violated scalar perturbations, that originates during inflation \cite{ha_ref_1}, must imprint a corresponding signature on the LSS and also on $E$-mode, which in turn can induce CHA signature at small angular scales in $B$ modes due to weak lensing \footnote{SI violation in tensor sector also retains its imprint in the $B$-mode power spectrum at low $l$ \cite{gw_modulated}.}. Such a SI violation signal would be a clinching irrefutable evidence for primordial origin of the CHA. This also naturally provides a handle on the corresponding CHA in LSS, which can be confirmed by the upcoming mission EUCLID \cite{euclid}. We estimate the induced SI violation signal in the $B$-mode polarization due to CHA in lensing potential ($\Phi$) and $E$-mode polarization. Finally we also estimate the measurability of the effect from future experiments.

The observed hemispherical asymmetry in CMB can be modelled by a dipole modulation in the scalar ($s$) and tensor ($t$) perturbations as \cite{gordon} 
\begin{equation}\label{eq1}
\tilde{\mathcal{X}}(\hat n) = (1+ \,\alpha_{s,t} \,\hat p. \hat n)\mathcal{X}(\hat n),
\end{equation}
where $\mathcal{X}= T, E, B, \Phi$ and $\alpha_{s,t}$ is the modulation strength for scalar \& tensor perturbations. The direction of dipole modulation $\hat{p}$ is denoted by $(l,b) = (228^\circ,-18^\circ)$ as measured by Planck. CHA or any other kind of SI violation can be measured using  Bipolar Spherical Harmonics (BipoSH) coefficients $A^{JM}_{ll'}$ introduced  by Hajian \& Souradeep \cite{ts}. These are related to the off-diagonal terms of the covariance matrix $\langle \mathcal{X}_{lm} \mathcal{X}^{'}_{l'm'}\rangle$  of the random field $\mathcal{X}(\hat n)$ on the sphere as 
\begin{align}\label{eq2}
\begin{split}
\big\langle \tilde{\mathcal{X}}_{lm} \tilde{\mathcal{X}}^{*'}_{l'm'}\big\rangle = \sum_{JK} A^{JK}_{ll'|\mathcal{X}\mathcal{X}^{'}}(-1)^{-m'}C^{JK}_{lm l' -m'} ,
\end{split}
\end{align}
where, $C^{JK}_{lml'm'}$ are the Clebsch-Gordan coefficients and $ \mathcal{X}_{lm} = \int d^2\hat n\, \mathcal{X}(\hat n)Y^*_{lm}(\hat n)$. The dipole model considered  in Eq. \ref{eq1}, leads to non-zero value of BipoSH coefficients for $J=1$. These BipoSH coefficients are also related to the local angular power spectra $C_l( \hat n)$ at a direction $\hat n$ \cite{yashar, muk_dirn_param}. 

 The polarization field ${}_{\pm2}X(\hat n)= Q(\hat n) \,\pm \,i\,U(\hat n)$ of CMB can be described as \cite{pol}
\begin{align}\label{eq3}
\begin{split}
{}_{\pm2}X(\hat n)= &\sum_{lm} {}_{\pm2}X_{lm} \,{}_{\pm2}Y_{lm}(\hat n),\\
{}_{\pm2}X(\hat n)=& \sum_{lm} (E_{lm} \pm i B_{lm}) \,{}_{\pm2}Y_{lm}(\hat n),
\end{split}
\end{align}
where, $Q$ and $U$ are the Stokes parameters. The induced CHA on polarization field ${}_{\pm2}X(\hat n)$ can be completely expressed  as 
 \begin{widetext}
\begin{align}\label{eqbi2a}
\begin{split}
{}_{\pm2}\tilde{X}(\hat n)  &= (1+ \alpha_{s,t} \, \hat p.\hat n){}_{\pm2}X(\hat n) + \bigtriangledown_{i} ((1+ \alpha_{s,t} \, \hat p.\hat n) \Phi (\hat n))  \bigtriangledown^{i} ((1+ \alpha_{s,t} \, \hat p.\hat n) {}_{\pm2}X(\hat n)) +\\ & \frac{1}{2} \bigg[\bigtriangledown_{i} ((1+ \alpha_{s,t} \, \hat p.\hat n) \Phi (\hat n)) \bigtriangledown_{j} ((1+ \alpha_{s,t} \, \hat p.\hat n) \Phi (\hat n)) \bigg]\bigg[\bigtriangledown^{i} \bigtriangledown^{j} ((1+ \alpha_{s,t} \, \hat p.\hat n) {}_{\pm2}X(\hat n))\bigg].
\end{split}
\end{align}
To capture the complete effect, we consider the lensing due to SI violated lensing potential $\tilde {\Phi}(\hat n)$ on SI violated polarization field ${}_{\pm2}\tilde X(\hat n)$. In spherical harmonic basis, Eq. \ref{eqbi2a} can be expressed as,
\begin{align}\label{eqbi3a}
\begin{split}
 {}_{\pm2}\tilde{X}_{lm} = & {}_{\pm2}{X}_{lm} + 
 \sum_{\substack{{l_1m_1}\\ {l_2m_2}}} \Phi_{l_1m_1}{}_{\pm2}{X}_{l_2m_2} \bigg[{}_{\pm2}H_{ll_1l_2}^{mm_1m_2}  + 
\sum_{l_3m_3} \Phi^{*}_{l_3m_3}{}_{\pm2}I_{ll_1l_3l_2}^{mm_1m_3m_2}\bigg]  +
 \sum_{\substack{K\\ {l_1m_1}}} \alpha^{1K}_{s,t} {}_{\pm2}{X}_{l_1m_1} {}_{\pm2}G_{ll_11}^{mm_1K}  \\&
+ \sum_{\substack{{K}\\{l_1m_1}\\ {l_2m_2}\\ {JN}}}\alpha^{1K}_{s,t} \Phi_{l_1m_1} {}_{\pm2}{X}_{l_2m_2}\bigg[ {}_{\pm2}H_{lJl_2}^{mNm_2}  {}_{0}R^{NKm_1}_{J1l_1} + {}_{\pm2}H_{ll_1J}^{mm_1N}  {}_{\pm2}R^{NKm_2}_{J1l_2}\bigg] \\&
  + \sum_{\substack{{K}\\{l_1m_1}\\ {l_2m_2}\\{l_3m_3}\\{JN}}}\alpha^{1K}_{s,t} \Phi_{l_1m_1}\Phi^{*}_{l_2m_2}{}_{\pm2}{X}_{l_3m_3}\bigg[{}_{\pm2}I_{lJl_2l_3}^{mNm_2m_3}{}_{0}R^{NKm_1}_{J1l_1}  + 
    {}_{\pm2}I_{ll_1Jl_3}^{mm_1Nm_3} {}_{0}R^{NKm_2}_{J1l_2} + {}_{\pm2}I_{ll_1l_2J}^{mm_1m_2N} {}_{\pm2}R^{NKm_3}_{J1l_3}\bigg], \\
\text{where,} \\ {}_{\pm2} G_{ll_1l_2}^{mm_1m_2}=& \int d^2 \, \hat n\, \bigg({}_{\pm2}Y_{l_1m_1}(\hat n) Y_{l_1m_1}(\hat n)\, {}_{\pm2}Y^{*}_{lm}(\hat n)\bigg); \,\,\,\,\,\,\,
{}_{\pm2} H_{ll_1l_2}^{mm_1m_2}= \int d^2 \hat n \bigg( \bigtriangledown_{i} Y_{l_1m_1}(\hat n) \bigtriangledown ^{i} {}_{\pm2}Y_{l_2m_2}(\hat n)\, {}_{\pm2}Y^{*} _{lm}(\hat n)\bigg), \\
{}_{\pm2} I_{ll_1l_2l_3}^{mm_1m_2m_3}=&  \frac{1}{2}\int d^2 \hat n \bigg(\bigtriangledown^{i} Y_{l_1m_1}(\hat n) \bigtriangledown^{i}  Y_{l_2m_2}(\hat n)\,
\bigtriangledown _{i}\bigtriangledown _{j} {}_{\pm2}Y_{l_2m_2}(\hat n) \,{}_{\pm2}Y^{*} _{lm}(\hat n)\bigg);\,\,\,\,
{}_{\pm s}R^{mm_1m_2}_{ll_1l_2}= \frac{\Pi_{l_1l_2}}{\sqrt{4\pi} \Pi_{l}}C^{l\,\pm s}_{l_1\,0\,l_2\,\pm s}C^{lm}_{l_1\,m_1\,l_2\,m_2}. 
\end{split}
\end{align}
Here, the first three terms are the usual contribution for SI polarization field and the last two terms arises due to SI violation. On assuming no primordial tensor perturbations (tensor to scalar ratio $r=0$)\footnote{Current observational bound on $r$ is consistent with zero \cite{planck_r}.}, the off-diagonal terms of two point correlation function originating due to SI violation in the scalar sector can be written as
\begin{align}\label{eqbi4}
\begin{split}
\bigg\langle {}_{\pm 2}\tilde X_{lm} {}_{\pm 2}\tilde X^{*}_{l'm'} \bigg\rangle = & \sum_{K} \bigg[\alpha^{1K} C_{l'}^{EE} {}_{\pm2}G_{ll'1}^{mm'K}  + \alpha^{*1K} C_{l}^{EE} {}_{\pm2}G_{l'l1}^{*m'mK} \bigg]  \\ & 
+\sum_{\substack{{Jl_1l_2}\\{KNm_1m_2}}}C_{l_1}^{\Phi\Phi}C_{l_2}^{EE}  \bigg[\alpha^{1K} {}_{\pm2} H_{l'l_1l_2}^{*m'm_1m_2} \bigg( {}_{\pm2}H_{lJl_2}^{mNm_2}{}_{0}R^{NKm_1}_{J1l_1} + {}_{\pm2}H_{ll_1J}^{mm_1N}{}_{\pm2}R^{NKm_2}_{J1l_2} \bigg) \\ &
+  \alpha^{*1K}{}_{\pm2}H_{ll_1l_2}^{mm_1m_2}  \bigg({}_{\pm2}H_{l'Jl_2}^{*m'Nm_2}{}_{0}R^{*NKm_1}_{J1l_1} + {}_{\pm2} H_{l'l_1J}^{*m'm_1N}{}_{\pm2}R^{*NKm_2}_{J1l_2} \bigg)\bigg] \\ &
+ \sum_{\substack{{Jl_1}\\{KNm_1}}}C_{l_1}^{\Phi\Phi} \bigg[\alpha^{1K} C_{l'}^{EE} \bigg( {}_{\pm2}I_{lJl_1l'}^{mNm_1m'}{}_{0}R^{NKm_1}_{J1l_1} 
+  {}_{\pm2}I_{ll_1Jl'}^{mm_1Nm'} {}_{0}R^{NKm_1}_{J1l_1} 
+  {}_{\pm2}I_{ll_1l_1J}^{mm_1m_1N} {}_{\pm2}R^{NKm'}_{J1l'}\bigg)\\& + \alpha^{*1K} C_{l}^{EE} \bigg( {}_{\pm2}I_{l'Jl_1l}^{*m'Nm_1m}{}_{0}R^{*NKm_1}_{J1l_1} 
+  {}_{\pm2}I_{l'l_1Jl}^{*m'm_1Nm} {}_{0}R^{*NKm_1}_{J1l_1} 
+  {}_{\pm2}I_{l'l_1l_1J}^{*m'm_1m_1N} {}_{\pm2}R^{*NKm}_{J1l}\bigg)\bigg]  \\ &+ \sum_{\substack{{Jl_1l_2}\\{KNm_1m_2}}}C_{l_1}^{\Phi\Phi} C_{l_2}^{EE} \bigg[\alpha^{1K}  \bigg( {}_{\pm2}I_{l'l_1l_1l_2}^{m'm_1m_1m_2}{}_{\pm2}G_{ll_21}^{mm_2K} \bigg) + \alpha^{*1K}  \bigg( {}_{\pm2}I_{ll_1l_1l_2}^{mm_1m_1m_2}{}_{\pm 2}G_{l'l_21}^{m'm_2K} \bigg)\bigg].
\end{split}
\end{align}
\end{widetext}
Theoretically, the inclusion of non-zero $r$ within the bounds would not alter the primary conclusions of this work. Unlike $T$ \& $E$, the secondary contribution to $B$-mode polarization due to weak lensing is dominant over the primary contribution that originates at the surface of last scattering. As a result, the signature of SI violation originating from weak lensing  is also most dominant on $B$ modes and hence provides best window for measuring this effect. So we explicitly calculate the contribution only for $B$-mode polarization. However, these calculations can also be readily extended to $T$ \& $E$. Since\textcolor{red}{,} Planck \cite{planck23, planckis,planck-oth} has measured a scale dependent SI violation, we also incorporate the same on the modulation field by taking $\alpha^{1K} \equiv \alpha^{1K}_{l}$. 
The analytical expression of BipoSH coefficients for $B$-mode polarization due to the scale dependent modulation field is
\begin{widetext}
\begin{align}\label{eqbi5a}
\begin{split}
 A_{ll+1|BB}^{10}=  & \sum_{Jl_1l_2}^{(l_1)_{max}} \frac{\alpha^{10}_{l_1}}{2\sqrt{4\pi}}\bigg[C_{l_1}^{\Phi\Phi}\big[C_{l_2}^{EE}- (-1)^{l+1+l_1+l_2}C_{l_2}^{EE}\big]\bigg[M_{Jl_2l}M_{l_1l_2l+1} C^{J0}_{10l_10}C^{l\,2}_{J0l_2\,2}C^{l+1\,2}_{l_10l_2\,2} \Pi_{l_1Jl_2l_2}\mathcal{W}_{l_1l_2l+1J1l}\bigg] + \\& C_{l_1}^{\Phi\Phi}\big[C_{l_2}^{EE}- (-1)^{l+1+J+l_2}C_{l_2}^{EE}\big]\bigg[M_{l_1l_2l}M_{Jl_2l+1} C^{J0}_{10l_10}C^{l\,2}_{l_10l_2\,2}C^{l+1\,2}_{J0l_2\,2} \Pi_{l_1Jl_2l_2}\mathcal{W}_{l_1l_2lJ1l+1}\bigg] \bigg]+\\& 
\sum_{Jl_1l_2}^{(l_2)_{max}}  \frac{\alpha^{10}_{l_2}}{2\sqrt{4\pi}} \bigg[C_{l_1}^{\Phi\Phi}\big[C_{l_2}^{EE}- (-1)^{l+1+l_1+l_2}C_{l_2}^{EE}\big]\bigg[M_{l_1Jl}M_{l_1l_2l+1} C^{J2}_{10l_22}C^{l\,2}_{l_10J\,2}C^{l+1\,2}_{l_10l_2\,2} \Pi_{l_1l_1Jl_2}\mathcal{W}_{l_2l_1l+1J1l}\bigg] +\\& C_{l_1}^{\Phi\Phi}\big[C_{l_2}^{EE}- (-1)^{l+1+l_1+J}C_{l_2}^{EE}\big]\bigg[M_{l_1l_2l}M_{l_1Jl+1} C^{J2}_{10l_22}C^{l\,2}_{l_10l_2\,2}C^{l+1\,2}_{l_10J\,2} \Pi_{l_1l_1Jl_2}\mathcal{W}_{l_2l_1lJ1l+1}\bigg] \bigg], \\
\,
A_{ll+1|BB}^{10}=& \sum_{l_1}^{(l_1)_{max}} \alpha^{10}_{l_1} S^{10}_{ll+1l_1},\\
 \text{where,}\,\,
\mathcal{W}_{l_2l_1lJ1l'}=& \begin{pmatrix}
    l_2       & l_1 & l' \\
    l       & 1 & J \\
\end{pmatrix};\,
M_{l_1l_2l}= \frac{1}{2\sqrt{4\pi}}[l_1(l_1+1)+l_2(l_2+1) - l(l+1)];\,
\Pi_{l_1l_2\dots l_n}= \sqrt{(2l_1+1)(2l_2+1)\dots (2l_n+1)}.
\end{split}
\end{align}
\end{widetext}
Here, $S^{10}_{ll+1l_1}$ is the shape factor that includes all the dependencies arising from weak lensing and $\mathcal{W}_{l_2l_1lJ1l'}$ denotes the Wigner $6j$-symbol. With no loss of generality, we also assume coordinates such that  the direction of hemispherical asymmetry is along the $z$ direction which leads to non-zero BipoSH coefficients $A^{JK}_{ll'}$ only for $K=0$. To obtain Eq. \ref{eqbi5a}, we used the following relations of spherical harmonics \cite{wayne, varsha},
\begin{align}\label{eqprop}
\begin{split}
{}_{\pm2} G_{ll_1l_2}^{mm_1m_2}= & \frac{\Pi_{l_1\, l_2}}{\sqrt{4\pi}\Pi_{l}}C^{l \pm 2}_{l_1\,0\, l_2 \,\pm 2}C^{lm}_{l_1\, m_1 \,l_2\, m_2},\\
{}_{\pm2} H_{ll_1l_2}^{mm_1m_2}=& \frac{1}{2}\frac{\Pi_{l_1\, l_2}}{\sqrt{4\pi}\Pi_{l}} \bigg[l_2(l_2+1)+l_1(l_1+1)-l(l+1)\bigg] \\& \bigg(C^{l \pm2}_{l_1\,0\, l_2\, \pm 2}C^{l m}_{l_1\, m_1\, l_2\, m_2}\bigg).
\end{split}
\end{align}

The scale dependent nature of modulation strength leads to only a limited contribution (up to certain $l_{max}$) from $C_{l}^{\Phi\Phi}$ and $C_{l}^{EE}$. As a result, different scale dependencies of $\alpha^{10}_{l}$ produce   different features in BipoSH coefficients. This makes it a decisive observable  to ascertain the origin including the scale dependence of CHA. A different modulation field for $\Phi$ and $E$-mode can be incorporated by taking different values of $\alpha^{10}_{l}$ in Eq. \ref{eqbi5a}. The value of $\alpha^{10}_{l}$ can also be considered zero for either of $\Phi$ or $E$-mode for comparing with the measurements. Due to the fact that the lensing contribution from matter occurs at much lower redshift in comparison to the surface of last scattering ($z\approx 1100$), we expect the effect of CHA to be present in $\Phi$ at much large angular scales than in $E$ and $T$. Hence\textcolor{red}{,} there are very few modes to measure the effect of CHA from lensing potential $\Phi$.

\begin{figure}[h]
\centering
\subfigure{
\includegraphics[width=3.8in,keepaspectratio=true]{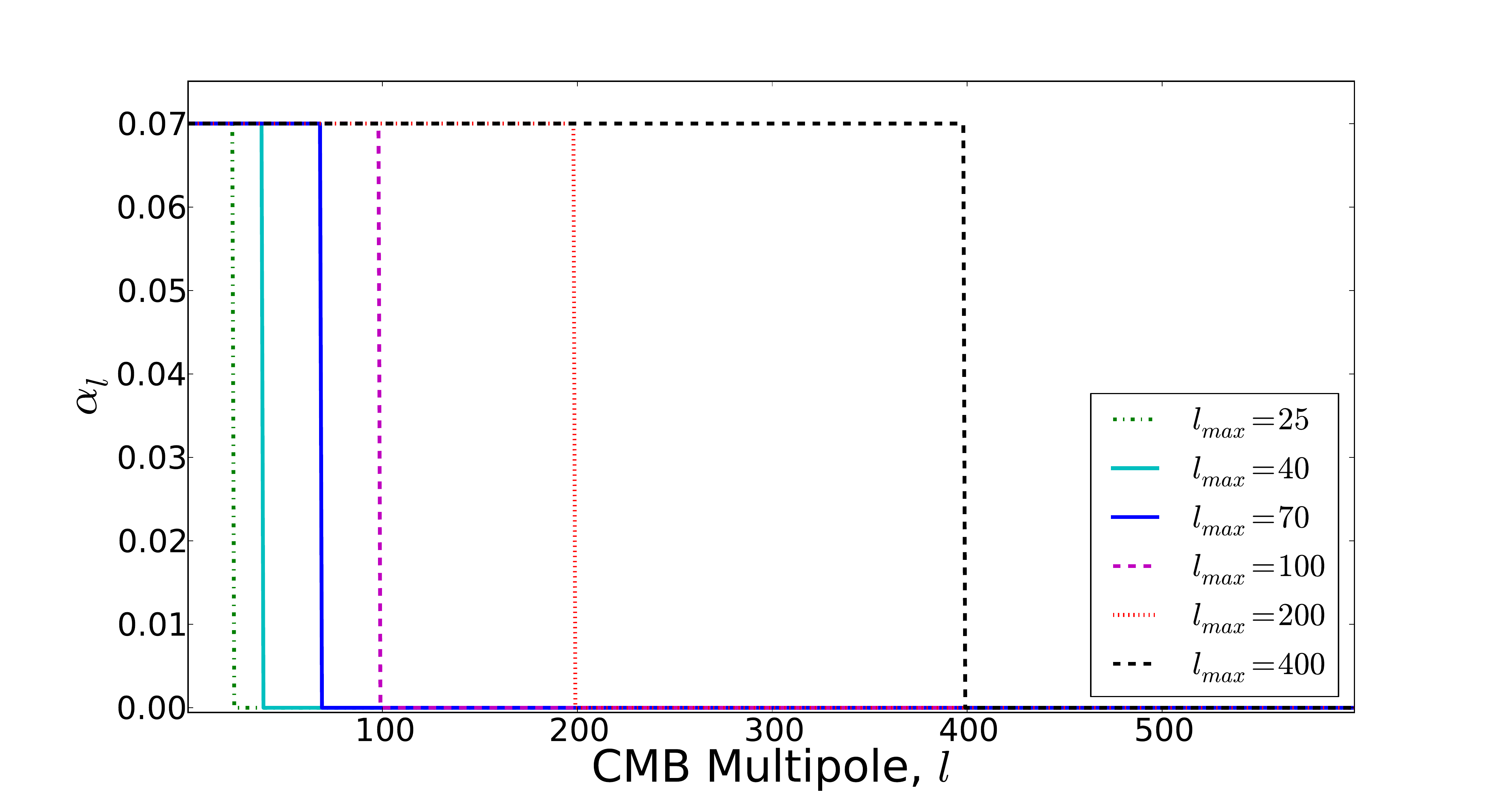}\label{fig1a}
}
\subfigure{
\includegraphics[width=3.8in,keepaspectratio=true]{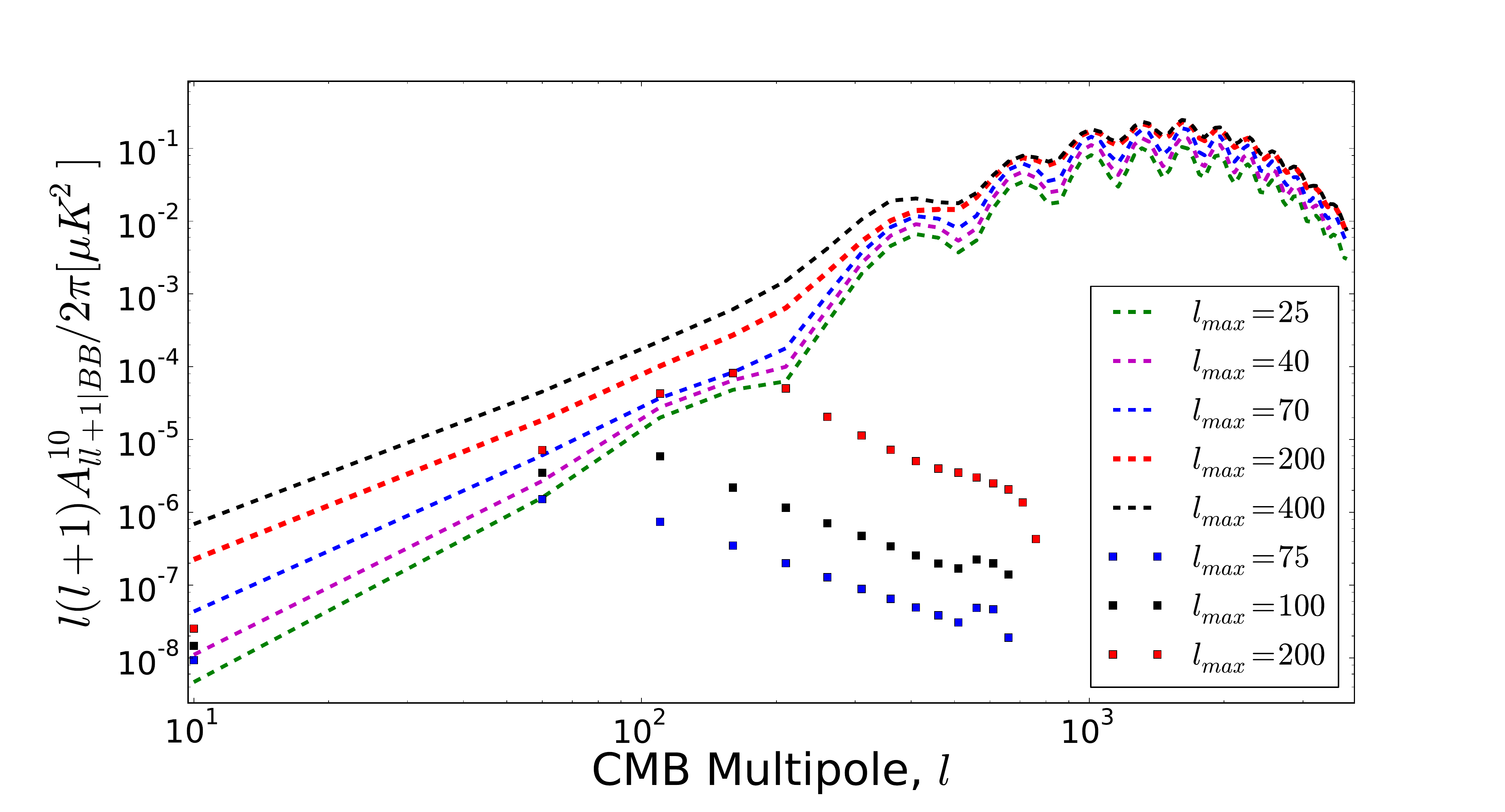}\label{fig1}
}
\captionsetup{singlelinecheck=off,justification=raggedright}
\caption{(a) We plot six different types of scale dependent modulation strength $\alpha_{l}$ with spatial extent up to $l_{max} =25, \,40,\,70,\, 100,\,200$ and $400$ which are considered in our analysis. (b)BipoSH coefficients for $B$-mode polarization that arise due to joint contributions from dipole modulated scalar perturbation in both lensing potential $\Phi$ and $E$ modes  are plotted in dashed lines. The contribution only from $E$ mode polarization are plotted by squares. }
\end{figure}

To understand the implication of CHA on $B$ modes due $\Phi$ and $E$ mode polarization,  we consider a few scale dependent cases of modulation field with $\alpha_{l} =0.07$ as depicted in Fig. \ref{fig1a}. With the same $\alpha_{l} =0.07$ for both $E$ and $\Phi$, we calculate the BipoSH coefficients for six different cases of scale dependent modulation amplitude as shown in Fig \ref{fig1a}. The numerical value of the BipoSH coefficients depicted in Fig. \ref{fig1}, are obtained using $C_l^{\Phi\Phi}$ \& $C_l^{EE}$ from CAMB \cite{camb} with the best fit cosmological parameters \cite{planck15_param}. As shown in Fig. \ref{fig1}, the presence of CHA at large angular scales in $\Phi$ and $E$-mode polarization generate CHA signatures in $B$ modes at small angular scales ($l\approx 1000$). This happens due to mixing of power from $E$ modes to $B$ modes at small angular scales.  Primarily, the amplitude of $A^{10}_{ll+1|BB}$ increases and secondly, the oscillatory behaviour at large $l$, gets smeared as the modulation field extends to smaller scales. Both these features appear due to increase in the contributions from $C_l^{\Phi\Phi}$ \& $C_l^{EE}$ in Eq.\ref{eqbi5a}.  In the direction of CHA ($\hat p$), the contribution from modulation field is maximum, which results in a larger value of $C_l^{BB}$ at small scales than the average value of $C_l^{BB}$. Whereas in the direction opposite to CHA ($-\hat p$), $C_l^{BB}$ is lower than the average value.  To estimate the effect only from $E$ mode polarization, we calculate Eq. \ref{eqbi5a} with non-zero values of $\alpha_l$ only for $E$ modes. As depicted in Fig. \ref{fig1}, the contribution to $B$ mode BipoSH coefficients from $E$ mode polarization is significantly weak in comparison to the joint contribution from both $\Phi$ and $E$ modes. This is due to two facts, firstly  the number of modes which contributes to Eq. \ref{eqbi5a} is much lower and secondly the contribution from lensing at small scales is negligible. This implies that the major contribution in the BipoSH signal arises from lensing potential $\Phi$ and not from $E$ mode polarization.  This gives us a new  insight into CHA in generating SI violated $B$ mode polarization at small scales.  

\begin{table*}
\centering
\caption{Expected cumulative noise $\sigma_N$ and SNR for measurement of BipoSH signal from lensing potential. Here we considered $l_{min}=2$ in Eq. \ref{noise1}.}
\label{tab_1} 
\vspace{0.5cm}
\begin{tabular}{|p{2.5cm}|p{2.5cm}|p{2.5cm}|}
\hline 
\centering Extent of CHA in lensing $(l_{max})$ & \centering cumulative noise $\sigma_N$ & \centering Signal to Noise Ratio (SNR) \tabularnewline
\hline
\centering $25$ & \centering $0.054$ &  \centering$1.28$ \tabularnewline
\hline
\centering $30$ & \centering $0.0456$ & \centering$1.53$ \tabularnewline
\hline
\centering $40$ & \centering $0.034$ & \centering$2.03$ \tabularnewline
\hline
\centering $70$ & \centering $0.02$ & \centering$3.46$ \tabularnewline
\hline
\end{tabular}
\end{table*}

This new avenue to measure the imprint of CHA in the lensing field $\Phi$ is a very important probe to measure the scale dependence of the asymmetry in the matter distribution. The scale dependence of CHA in the matter sector has been investigated earlier using quasar number count by Hirata \cite{hirata}, showing that the amplitude of power asymmetry is less than $1\%$ for the wavenumber $k = 1.3-1.8\, h\, \text{Mpc}^{-1}$. More recently Flender \& Hotchkiss \cite{flender} imposed a stronger bound on the scale dependence of CHA. Using the Planck's SMICA map, they showed that the amplitude of the asymmetry is below $0.0045$ at $95\%$ confidence limit for $k\approx 0.06-0.2\, h\,\text{Mpc}^{-1}$. These two measurements make it more viable for the CHA to be present at $7\%$ level for k $<0.06\,h\, \text{Mpc}^{-1}$.  The effect mentioned in this Letter can probe a smaller $k$ range of the matter distribution (low $l$ of lensing potential $C_l^{\Phi\Phi}$), which is not probed by earlier studies \cite{hirata, flender}. The signatures in $B$ mode BipoSH spectra ($A^{10}_{ll+1|BB}$) which peak around $l \sim 1000$ gives an independent bound on the amplitude and scale dependence of the modulation field.

Since the effect of CHA in lensing potential contributes maximum to $B$ mode BipoSH coefficients, we expect the measurability of this phenomenon to depend upon associated noise of each independent modes in $\Phi$ which carries the information about CHA. Each mode, $l$ of $C_l^{\Phi\Phi}$ have an associated noise of $\sqrt{\frac{2}{2l+1}}C_l^{\Phi\Phi}$. So to estimate the relative change in amplitude ($\alpha = \frac{\Delta C_l}{C_l}$) due to CHA in lensing potential up to $l= l_{max}$, we construct the total associated cumulative noise ($\sigma_N$) for the $l$ range [$l_{min}$,\,$l_{max}$] as\footnote{This is also true for estimation of CHA amplitude from temperature field. The error bar of the BipoSH estimator for first $64$ independent modes is $0.022$ \cite{planckis}, which can also be estimated by  similar approach.}
\begin{equation}\label{noise1}
\sigma_N= \sqrt{\frac{1}{\sum_{l=l_{min}}^{l_{max}}\frac{2l+1}{2}}}.
\end{equation}
So, if CHA is present with constant amplitude $\alpha$ up to $l=l_{max}$ in lensing potential, then it can be measured at-best with a noise $\sigma_N$ from lensing potential. This translates into the fact that the maximum significance with which modulation amplitude $\alpha$ can be estimated from $B$ mode BipoSH coefficients is also limited by $\sigma_N$. The signature of CHA which is present in each mode of lensing potential gets divided into several correlated modes of $B$ modes polarization. Also because of the fact that CHA from $E$ modes make negligible contribution to the $B$ mode BipoSH coefficients, the main detectability of the signal depends on the variance from $\Phi$ and not from $E$ modes. In Table \ref{tab_1}, we mention the $\sigma_N$ and corresponding SNR for different $l_{max}$ values of lensing potential. Since, the exact scale dependence of CHA in $T$ and $\Phi$ is not yet known, we consider $4$ different cases of scale dependence. For non-zero modulation amplitude up to $l<25$, the $B$-mode BipoSH coefficients cannot be measured with high significance. Whereas, a non-zero modulation amplitude up to $l = 70$, can lead to more than $3\sigma$ detection. Stage-IV CMB polarization experiment (CMB-S4) \cite{cmbs4} with large number of detectors ($10^5-10^6$), small beam size and large sky coverage ($f_{sky}$) can reach an unprecedented instrumental noise $N_p \leq 1$ $\mu$K-arcmin \cite{cmbs4}. This allows to measure the $B$ mode polarization to small angular scales ($l \sim 2000$). Such an exquisite measurement of $B$ mode polarization enables to detect the BipoSH signal $A^{10}_{ll+1|BB}$ (which peaks around $l \sim 1000$) with an accuracy limited by cosmic variance. However, future experiment with sky coverage $f_{sky}$ leads to $1/\sqrt{f_{sky}}$ degradation of SNR than quoted in Table \ref{tab_1}. With already available detector technology, measurement of this signal is feasible within the next few years.  Within the framework of current cosmological models \cite{ha_ref_1} that explains CHA, the exact scale dependence of the CHA is neither completely understood nor validated by observations. If future missions measure power asymmetry in $B$ modes at small scales, then that can be attributed to CHA in lensing potential and hence can be an independent window to understand the spatial extent and also the origin of this asymmetry.  A detailed expression
for the covariance matrix of BipoSH coefficients is given in Appendix A.

In this Letter, we show that CHA arising from scalar sector of cosmic perturbations can lead to CHA in CMB $B$-mode polarization at small angular scales. $B$-mode polarization gets significantly enhanced at small angular scales ($l \approx 1000$) due to inevitable leakage of power from scalar perturbations through weak lensing. As a consequence, it captures CHA from lensing potential $\Phi$ and $E$ modes and shows significant non-zero BipoSH coefficients at large $l$, in contrast to $T$ \& $E$-mode at low $l$. The contribution to BipoSH coefficients dominantly arise from the CHA in lensing potential $\Phi$ and not from $E$ modes as shown in Fig.\ref{fig1}. The signatures of CHA is obscured by cosmic variance at low $l$ for $T,\,E$-mode \& $\Phi$, and its range of exact spatial scales is difficult to unravel. The effect brought to light in this letter can provide an independent window to estimate the spatial extent of CHA in lensing potential. For the scale dependent CHA in scalar perturbations which decays by $l \approx 70$ in temperature and polarization, affect $\Phi$ at much lower multipoles ($l<25$). As a result, the effect of this from $B$ mode cannot be detected due to large contribution from cosmic variance. However, if the spatial extent of CHA is present up to $l \approx 40$ in $\Phi$, then a mild $2\sigma$ detection of this signal is possible as indicated in Table \ref{tab_1}. The physical mechanism mentioned here is an independent window to constraint the scale dependence of CHA. Detection of dipolar power asymmetry in $B$ mode polarization from small angular scales can improve our understanding on the spatial range of scales affected by this anomaly.
 
\textbf{Acknowledgements} S.~Mukherjee acknowledges Council of Scientific \& Industrial Research (CSIR), India for financial support as Senior Research Fellow. We thank James P. Zibin for useful comments. The computational work is carried out on the High Performance Computing facility at IUCAA.

\section{Appendix A: Covariance matrix of BipoSH coefficients for B mode polarization}\label{app}
The estimation of the BipoSH signal from the observation and to estimate the corresponding modulation amplitude $\alpha$, it is important to derive the expression for the covariance matrix of BipoSH coefficients of the $B$ mode polarization. The correlation between the CMB multipoles, $l$ in $B$ mode polarization \cite{cooray, hu} leads to non-zero off-diagonal terms in the covariance matrix of $B$ mode BipoSH coefficients. The elements of covariance matrix for $B$ mode BipoSH coefficients are,
\begin{widetext}
\begin{align}
\begin{split}
\langle A^{1M*}_{l_1l_2|BB}A^{1M'}_{l_3l_4|BB}\rangle &=  C_{l_1}^{BB}C_{l_3}^{BB}(-1)^{l_1+l_2+1}\delta_{MM'}\delta_{l_1l_4}\delta_{l_2l_3} + C_{l_1}^{BB}C_{l_2}^{BB}\delta_{MM'}\delta_{l_1l_3}\delta_{l_2l_4} + 
\\& \frac{1}{16}\bigg[\sum_{l_5\,l_6\,l_7\,l_8} C^{\Phi\Phi}_{l_5}C^{\Phi\Phi}_{l_6}C^{EE}_{l_7}C^{EE}_{l_8}\bigg(\bigg\{\bigg((1-(-1)^{l_1+l_5+l_7})(1-(-1)^{l_2+l_5+l_8})(1-(-1)^{l_3+l_6+l_7})(1-(-1)^{l_4+l_6+l_8})\bigg)\\& F_{l_1l_5l_7}F_{l_2l_5l_8}F_{l_3l_6l_7}F_{l_4l_6l_8}\Pi_{l_1l_2l_3l_4}\sum_{sp}\bigg((N_{l_8l_5l_2l_6l_7l_3l_4l_1s})\Pi_{ss11}\sum_{JN} C^{JN}_{1M1M'}C^{JN}_{spsp}N_{l_4l_1sl_3l_2s11J}\bigg)\bigg\} + 
\\ & \bigg\{\bigg((1-(-1)^{l_1+l_5+l_7})(1-(-1)^{l_2+l_5+l_8})(1-(-1)^{l_3+l_6+l_8})(1-(-1)^{l_4+l_6+l_7})\bigg)\\&F_{l_1l_5l_7}F_{l_2l_5l_8}F_{l_3l_6l_8}F_{l_4l_6l_7}\Pi_{l_1l_2l_3l_4}\sum_{sp}\bigg((N_{l_7l_5l_1l_6l_8l_3l_4l_2s})\Pi_{ss11}\sum_{JN} C^{JN}_{1M1M'}C^{JN}_{spsp}N_{l_1l_21l_3l_41ssJ}\bigg)\bigg\} +
\\& \bigg\{\bigg((1-(-1)^{l_1+l_5+l_7})(1-(-1)^{l_2+l_6+l_7})(1-(-1)^{l_3+l_6+l_8})(1-(-1)^{l_4+l_5+l_8})\bigg)\\&F_{l_1l_5l_7}F_{l_2l_6l_7}F_{l_3l_6l_8}F_{l_4l_5l_8}\Pi_{l_1l_2l_3l_4}\sum_{sp}\bigg(N_{l_7l_5l_1l_6l_8l_3l_2l_4s}\Pi_{ss11}\sum_{JN} C^{JN}_{1M1M'}C^{JN}_{spsp}N_{l_3l_41l_1l_21ssJ}\bigg)\bigg\}+ 
\\& \bigg\{\bigg((1-(-1)^{l_1+l_5+l_7})(1-(-1)^{l_2+l_6+l_7})(1-(-1)^{l_3+l_5+l_8})(1-(-1)^{l_4+l_6+l_8})\bigg)\\&F_{l_1l_5l_7}F_{l_2l_6l_7}F_{l_3l_5l_8}F_{l_4l_6l_8}\Pi_{l_1l_2l_3l_4}\sum_{sp}\bigg(N_{l_8l_5l_3l_6l_7l_2l_4l_1s}\Pi_{ss11}\sum_{JN} C^{JN}_{1M1M'}C^{JN}_{spsp}N_{l_3l_2sl_4l_1s11J}\bigg)\bigg\} +
\\& \bigg\{\bigg((1-(-1)^{l_1+l_5+l_7})(1-(-1)^{l_2+l_6+l_8})(1-(-1)^{l_3+l_5+l_8})(1-(-1)^{l_4+l_6+l_7})\bigg)\\&F_{l_1l_5l_7}F_{l_2l_6l_8}F_{l_3l_5l_8}F_{l_4l_6l_7}\Pi_{l_1l_2l_3l_4}\sum_{sp}\bigg(N_{l_8l_5l_3l_6l_7l_4l_2l_1s}\delta_{s1}\delta_{pM}\bigg)\bigg\} +
 \\& \bigg\{\bigg((1-(-1)^{l_1+l_5+l_7})(1-(-1)^{l_2+l_6+l_8})(1-(-1)^{l_3+l_6+l_7})(1-(-1)^{l_4+l_5+l_8})\bigg)\\&F_{l_1l_5l_7}F_{l_2l_6l_8}F_{l_3l_6l_7}F_{l_4l_5l_8}\Pi_{l_1l_2l_3l_4}\sum_{sp}\bigg(N_{l_8l_5l_4l_6l_7l_3l_2l_1s}\delta_{s1}\delta_{pM}\bigg)\bigg\}\bigg)\bigg],
\end{split}
\end{align}
\end{widetext}
where, $N_{l_1l_2l_3l_4l_5l_6l_7l_8l_9}$ is the Wigner $9j$-symbol usually written as  
\begin{align}
N_{l_1l_2l_3l_4l_5l_6l_7l_8l_9} =& \begin{pmatrix}
    l_1       & l_2 & l_3 \\
    l_4       & l_5 & l_6 \\
    l_7 	&l_8 & l_9
\end{pmatrix};\,
\end{align}
 and $F_{l_3l_2l_1}= M_{l_3l_2l_1}\Pi_{l_1l_2l_3}(-1)^{l_2+2l_1}C^{l_3-2}_{l_-2l_20}$. Here the first two terms contributes only to the diagonal of the covariance matrix.
\def\urlprefix{}
\def\url#1{}
\bibliography{references}
\bibliographystyle{apsrev}
\end{document}